# Implementation of Compute Intensive Algorithms on Software Configurable Processor

Ganesh[1*], Rodrigues Steevan[2], Niranjan, U.C.[2]

*1*. M. Tech student, NMAMIT, Nitte 574 110, India  *2*. Manipal Dot Net Pvt. Ltd, Manipal 576 104, India
* ganeshsprabhu@gmail.com

**Abstract**: *Software configurable processors (SCP) implement compute intensive applications very efficiently on the special on-chip configurable hardware. The SCP by Stretch Inc. converts the compute-heavy algorithms into custom instructions, called extension instructions (EI) which run on the on-chip logic. The Processor interleaves the EI's between regular instructions and the on-chip hardware executes the algorithm in parallel, accelerating the application. This results in a performance gain of more than order of magnitude over an un-accelerated processor.*

*This paper explains the implementation of two compute intensive algorithms on Stretch SCP, namely (i) colour space conversion and (ii) histogram equalisation. The repeated processing required by these algorithms is made easier by the SCP which allows packing of multiple pixels into a vector. The vector processing makes SCP achieve high throughput. Profiling an application identifies compute-intensive spots in the program, which are computed on the on-chip hardware by issuing EI's.*

## I. INTRODUCTION

The unique feature of software configurable processor [2] is that the compute intensive part of the application code can be run on the on-chip hardware known as ISEF. The most computationally expensive part in video/image processing is the repetitive pixel manipulation. By running these compute intensive programs on the SCP's ISEF, one can reduce the execution cycles, thereby speeding up the applications. The operands for the ISEF are passed with the help of wide register, which are 128 bits in width. This high bandwidth input, further worked upon by vector processing improves the speed of computation dramatically. Programming of the SCP to fit the application configures the on-chip hardware and this configuration takes the form of an extended instruction (EI). The processor issues appropriate EIs to run the application.

The colour space conversion algorithms are used for converting data from one colour space to another [1]. Typically this is necessitated in devices such as monitors and printers which work in different colour formats. The main goal of our colour space conversion algorithm running on SCP was to achieve high throughput by using vector processing on the ISEF. The colour system mainly uses set of equations to transform colour from one format to another. These equations are converted to fixed-point arithmetic to make them amenable for SCP operations.

Histogram equalisation [1], [5] is an image enhancement application and is known as contrast-stretching transformation. Low-contrast images result due to poor illumination, lack of dynamic range in the image sensors, and wrong setting of a lens aperture during image acquisition. The main idea behind contrast stretching is to increase the dynamic range of the Gray levels in low-contrast images. This image transformation requires computation of the image cumulative histogram and derives a look-up-table from it. Implementation of histogram equalisation on SCP increases the throughput in terms of pixel processing rate, when compared to a general-purpose processor implementation.

## II. SCP ARCHITECTURE

Software configurable processors are very useful for implementing compute intensive applications [2]. The S6000 is a family of SCP of Stretch Inc, whose architecture is shown in Fig. 1. The SCP has a core made up of Tensilica LX, which is VLIW RISC architecture. The software-configurable part of the processor is ISEF, which is capable of doing multiple arithmetic and logical operations in parallel. The ISEF consists of multiple multipliers, arithmetic units, registers and multiplexers; and is embedded in the processor. The S6000 for example is made up of 64 multiplication units each capable of 8x16bit multiplication, 4096 arithmetic units and 64KB of embedded RAM.

To speed up an application, one need to identify the parts of the program, which consume most compute-time. This is done using the profile utility and such compute-intensive parts of the code are called hot spots. The hot spots are computed on ISEF, which accelerate their execution. The operands for the ISEF are passed through the wide registers, which are 128bits wide. However, it is also possible to send the inputs to ISEF directly using DMA and store them in the ISEF RAM. The ISEF can pack and unpack the inputs sent through wide registers without consuming any cycles.

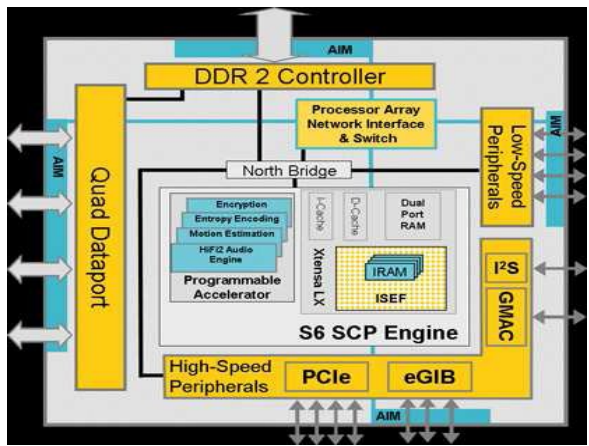

Fig. 1 The S6000 Family Architecture

The Stretch design tools enable the user to port the high-processor-cycle-consuming parts of the application code onto the ISEF, by writing them as Stretch extension functions [6] in C/C++ language. The Stretch extension functions running on the ISEF are known as extended instructions (EI). The execution of the EI in the ISEF with its inherent parallel processing accelerates the running of the application. The conversion of compute intensive part of the program into custom made processor instruction running on the on-chip hardware makes the SCP well suited for number of applications.

In addition, each S6000 member includes external memory support with a DDR2-667 SDRAM controller with 16- or 32-bit interface and an enhanced generic interface bus (GIB) for FLASH and other memory mapped peripherals. On-chip memory sub-systems include instruction and data cache, as well as a 64-KB block of SRAM. The 40 DMA controllers facilitate moving data on and off the devices with minimal processor interaction. The S6100 family member includes a four-lane PCIe interface. Other integrated peripherals include triple speed Ethernet MAC, two multi-channel inter–IC Sound (I2S) interfaces, two-wire interface (TWI), serial peripheral interface (SPI), two UARTs, and general purpose I/O (GPIO).

## III. IMPLEMENTATION OF COLOUR CONVERSION ALGORITHMS

Colour conversion [1], [8] algorithms convert colour attribute of a pixel/data from one colour space to another. Different colour space conversion algorithms are implemented on SCP, such as RGB to CMY, RGB to YCC, RGB to YUV, RGB to YIQ and RGB to CMYK[9]. The forward as well as reverse conversions are of hi-fidelity quality. Each of the colour conversion has different applications, for example the YIQ/YCC/YUV is television transmission primaries used extensively in TV Broadcasting and Receivers [3]. One of the conversion processes RGB to YIQ is explained in detail in the next section.

Fig. 2 depicts the general model for the colour conversion process on Software Configurable Processor. The colour of the input pixels is converted from one format to another. For real-time colour conversion of a QCIF video at 25 frames per second, the SCP should be able to process half mega pixels in a second.

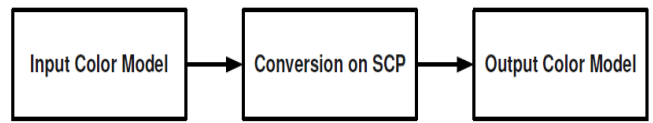

Fig. 2 Colour Conversion Model

### A. *RGB-YIQ Colour Space Conversion*

The following equations are used to convert pixel data from 8-bit RGB to 8-bit YIQ [1], [4]. As Stretch SCP is a fixed-point processor, these equations are converted into fixed-point arithmetic, by multiplying them with scaling factor of 256.

The conversion formula is,
  y = (77 * r + 150 * g + 29 * b) / 256;
  i = (153 * r - 70 * g - 82 * b) / 256;
  q = (54 * r - 134 * g + 80 * b) / 256;
The reverse conversion formula is
  r = (256 * y + 245 * i + 159 * q) / 256;
  g = (256 * y - 70 * i - 166 * q) / 256;
  b = (256 * y - 283 * i + 436 * q) / 256;

The three colour attributes of the pixels red, blue and green (r,g,b) are converted to the illumination and orthogonal chrominance components (y,i,q). The SCP used in this example runs at the clock frequency of 300 MHz.

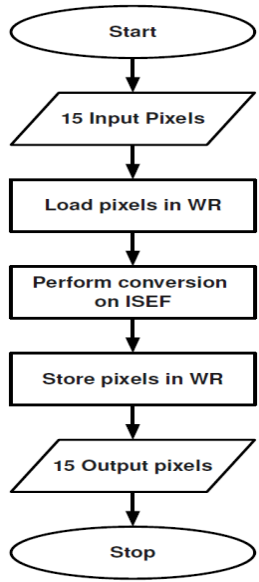

Fig. 3 Flowchart for Colour Space Conversion

Fig. 3 shows the step by step procedure to transform colour of pixels from one format to another on SCP.

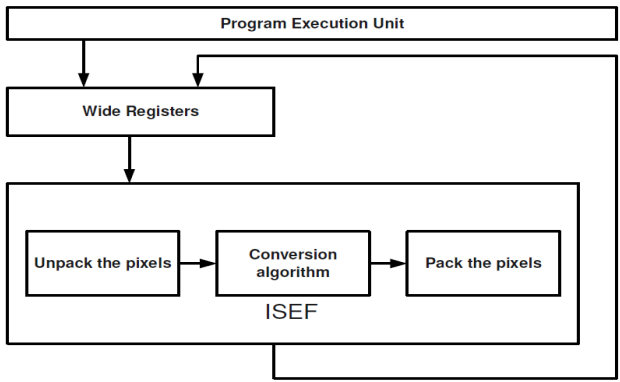

Fig. 4 Implementation of Colour Space Conversion

The performance of the code is measured through the utility profiling. Profiling the rgb-to-yiq function on the processor for 640 pixels but without using the ISEF resulted in 707524 cycles. This amounts to 11 process-cycles/pixel (707524/64000). It is observed that this result is consistent with that obtained by running the application on a board with SCP.

Then the rgb-to-yiq function is programmed as a custom instruction on the SCP, which accelerates the repetitive task of multiple pixel transformation. Pixel data are moved into and out of ISEF with the help of wide registers (WR), which are 128-bit long, as shown in Fig. 4. If one pixel is passed to the ISEF through the WR for every EI, then the total number of compute cycles is 234050. This is an improvement of more than 3x over the non-ISEF based implementation. It is also observed that only 30% of the available resources in ISEF were used for computation.

Wide registers are capable of holding 16 bytes and each pixel is stored in 3 bytes as (r,g,b) components. Thus, 5 pixels with 15 bytes were packed into a WR to make a vector and processed in parallel in the ISEF. This vector processing took 63518 cycles for colour conversion of 640 pixels. Although the available computational resources are not utilised completely, the full capacity of wide registers is made use of. The optimal use of the WR leads to reduced compute cycles.

The cycle measurement on the SCP board shows a few processor stalls, which increases the cycle count. Data cache miss is responsible for most of the stalls. Storing the input and transformed pixels in the internal data RAM whose latency is zero, eliminates the cache misses. It is also possible to use more than one wide register to improve the data bandwidth. By using two wide registers to move the data in and out of ISEF and with extended vector processing the cycle count was reduced to 72517(for 8 pixels).

The most optimised version of the application processes 640 pixels in 63518 cycles achieving a performance gain of 0.99 cycles/pixel. Also, compared with the non-ISEF based implementation, the application achieved a speed-up factor of 11 (707524 cycles/63518 cycles).

It is possible to convert different colour space system from one format to another [12], [13], by following the above flow. The flow initially involves identifying the compute-intensive parts [7] of the application and then programming the compute-intensive parts such that they are executed on the on-chip hardware. The programming part thus involves packing multiple data into WR and vector processing on ISEF [10],[11]. This programming configures the ISEF and which the processor sees as a custom-made extension instruction. Proper issue of extension instructions in the application program results in fast and efficient execution.

B. *ISEF Usage*

ISEF does not support the floating point arithmetic, due to limited on-chip computational resources. Therefore, the floating point operations are converted into equivalent fixed point arithmetic using scaling factors and LUTs (look up tables). All the built in operations in 'C' are not supported, but only a select few are supported. Trigonometric functions and division operations are not supported on ISEF. User has to write his own fixed point functions to mimic these unsupported operations.

Operations that cannot be easily converted into fixed-point arithmetic exist in RGB to HSI (HSV/HSL) conversion [13]. In such cases the computational parts requiring floating point are performed outside the ISEF. This still leads to enhanced performance (CMY-CMYK [12]). ISEF has special provision to pack and unpack the inputs sent through wide registers, which are processor cycle-free. Maximum of 3 wide registers can be used to pass the operands to the ISEF and 2 wide registers can be used to get the ISEF results.

C. *Running the Application*

The application program can be run and debugged on different targets such as native, remote and simulator. Native mode is faster and mainly used for quick-starting the application development. In this mode, the compiler builds the application for running on the x86 platforms (Windows or Linux). In the simulator mode the compiler simulates the behaviour of the SCP, with cycle-accurate precision. In the remote target, the applications are run on different boards comprising SCP and the generated ISEF configuration bit files are loaded on to the SCP.

## IV. HISTOGRAM EQUALISATION ON SCP

Histogram Equalisation [1] is an image processing technique widely used for contrast stretching of low-contrast images. The technique computes a pixel transformation function from the cumulative histogram of the given input image. In this example, the histogram and the transformation function are computed on the ISEF. The computed transformation function is stored in the IRAM (ISEF Random Access Memory) as a look up table, which is referenced for each pixel to generate enhanced-contrast image.

The Histogram Equalisation is implemented as a sequence of three Extension Instructions (EI). The first EI computes the histogram of 16 pixels simultaneously, storing the result in 16 banks of IRAM. As the SCP restricts use of one IRAM bank per EI, the histogram is broken up into sub histograms. The second EI merges the sub histograms to get cumulative histogram. The cumulative histogram acts as the pixel transformation function and is stored on 16 banks of IRAM as a look up table (LUT). The final EI accesses the LUT to transform Gray-scales of 16 pixels simultaneously. The use of multiple banks of IRAM with their simultaneous read-compute-write property accelerates the computations. The block diagram and the pictorial representation of the histogram equalisation technique are shown in Fig.5 and Fig.6 respectively.

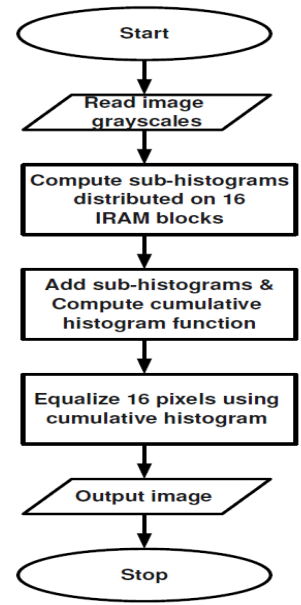

Fig. 5 Flowchart for Histogram Equalisation

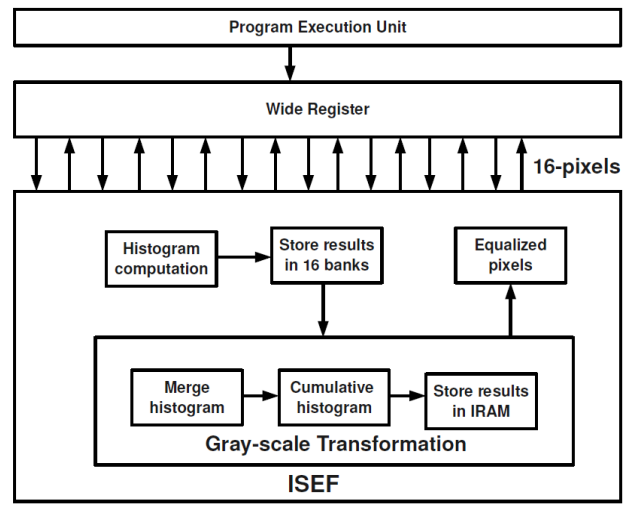

Fig. 6 Implementation of Histogram Equalisation

D. *Results*

This section reports the experimental results obtained from histogram equalisation for an image of size 128x128.

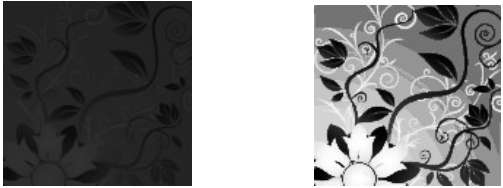

Fig. 7 Before histogram equalisation    Fig. 8 After histogram equalisation

Fig. 7 is the input low contrast image. The result of contrast stretching by the histogram equalisation technique is shown in Fig. 8. The cycle count for the non-ISEF based implementation of histogram equalisation was 17124334 cycles. By vector processing and optimal use of IRAM the compute cycles were reduced to 3154353.

E. *Observations of Histogram Equalisation*

The S6 processor stalls the issue of EIs during histogram computation when called repeatedly. This was seen in the pipeline trace, which shows that Xtensa pipeline is extended up to stage 28. However, wherever possible the compiler unrolls the loop comprising EI, by loading and unloading the pixels into independent set of WR registers and thereby reducing the processor stalls.

## V. CONCLUSIONS

We have presented the results of implementation of two image processing algorithms on the software configurable processor. The feature of the SCP namely, the running of the compute-intensive parts on the ISEF as extension instructions is exploited to enhance the throughput of the algorithms. The two algorithms are same in the sense that they transform an attribute of a pixel (colour and grey scale), but differ in the sense that one of them is implemented as a set of equations, while the other determines the transformation on the fly and implements it as a look-up-table. We achieved performance gain in compute-cycles per pixel of 0.99 and 1.9 for colour space conversion and histogram equalisation respectively. The corresponding acceleration factors for these two implementations are 11 and 5 over a non-ISEF based implementation with 46% and 12% utilisation of ISEF compute resource.

Stretch also provides a parallel processing platform called Software Configurable Processor Array (SCPA). This array is made of multiple SCPs, wherein one of them acts as master processor interacting with the external world. Stretch provides a very user-friendly set of APIs for programming the multiple processors. The application can be run as number of tasks running simultaneously on these processors. Different colour conversion algorithms such as RGB-YIQ and RGB-CMY can be run on the multiple SCP's in parallel. This would further speed up the application.

## ACKNOWLEDGMENT

We thank Stretch Inc USA and Manipal Dot Net Pvt. Ltd., for providing the opportunity to write this paper.